\documentclass[aps,prb,twocolumn,a4paper,showpacs]{revtex4-1}

\usepackage{amsmath}
\usepackage{amssymb}
\usepackage{color}
\usepackage{epsfig}
\usepackage{pgfplots}
\usepackage{scrextend}

\definecolor{green2}{rgb}{0.10000,0.450000,0.120000}%
\newcommand{\vecs}[1]{{\bf #1}}
\newcommand{\vphi}{\varphi}

\newcommand{\mrm}[1]{\mathrm{#1}}

\newcommand{\scst}[1]{{\scriptstyle #1}}

\begin{document}

\title{The phonon-assisted absorption of excitons in Cu$_2$O}
\author{Florian Sch\"one}
\author{Heinrich Stolz}
\affiliation{Institut f\"ur Physik, Universit\"at Rostock,
Albert-Einstein-Strasse 23, D-18059 Rostock, Germany}
\author{Nobuko Naka}
\affiliation{Department of Physics, Kyoto University, Kitshirakawa-Oiwake-cho, Sakyo-ku, Kyoto 606-8502, Japan}

\begin{abstract}
The basic theoretical foundation for the modelling of phonon-assisted absorption 
spectra in direct bandgap semiconductors, introduced by Elliott 60 years ago 
\cite{elliott1957intensity} using second order perturbation theory, results in 
a square root shaped dependency close to the absorption edge. 
A careful analysis of the experiments \cite{naka2005thin} reveals that for the 
yellow S excitons in Cu$_2$O the lineshape does not follow that square root 
dependence. The reexamination of the theory shows that the basic assumptions of 
constant matrix elements and constant energy denominators is invalid for 
semiconductors with dominant exciton effects like Cu$_2$O, where the phonon-assisted 
absorption proceeds via intermediate exciton states. The overlap between these and 
the final exciton states strongly determines the dependence of the absorption on 
the photon energy. To describe the experimental observed line shape of the indirect 
absorption of the yellow S exciton states we find it necessary to assume a momentum 
dependent deformation potential for the optical phonons. 
\end{abstract}


\maketitle



\section{Introduction}

The research focus in semiconductor physics has changed in
recent decades from generic bulk semiconductors in favor for physical phenomena in
more fancy systems, like lower dimensional structures. However, the discovery of
yellow excitons with principal quantum numbers up to $n=25$\cite{kazimierczuk2014giant} 
has renewed the interest of Cu$_2$O as it facilitates a novel branch of research 
in semiconductor physics 
\cite{grunwald2016signatures,thewes2015observation,schweiner2016linewidths,heckotter2017high,zielinska2016electro,schone2016deviations}. 
These highly excited states, generally referred to as Rydberg excitons, exhibit 
similar properties already observed in atom physics but in a much more experimentalist 
friendly framework (effects such as Rydberg blockade are
identifiable at liquid helium temperatures and the Stark effect manifests at rather low
electric field strengths \cite{heckotter2017high}). Furthermore, they additionally show new
characteristics due to the unique setting within the semiconductor \cite{schweiner2016linewidths,zielinska2016electro,schone2016deviations}. 
The cubic symmetry of the system leads e.g. to anisotropic band dispersions, fine-structure
splitting, or the breaking of antiunitary symmetries in magnetic fields \cite{schweiner2017magneto}.
Since the optical absorption bands of these Rydberg states sit on top of the phonon
assisted absorption into the yellow exciton ground state, exhibiting a strong Fano-
type interaction \cite{toyozawa1958theory}, a thorough understanding of the phonon-assisted
absorption processes is of uttermost importance for properties of the Rydberg
excitons.

The standard textbook approach to describe the shape of the phonon-assisted
exciton absorption close to the band gap is based on second-order perturbation theory
and goes back to Elliott \cite{elliott1957intensity}. It can be visualized as a direct optical
excitation into a dipole allowed virtual intermediate state and the subsequent
relaxation to the final state through the emission of a phonon. Then by assuming the
sum over the matrix elements and energy dominators to be constant, one can derive
the well-known square root dependence of the absorption coefficient \cite{kuper1963polarons}.
In this paper, we will critically examine these assumptions and show that in case of
semiconductors with strong exciton effects, like Cu$_2$O, they are invalid mainly due to
two reasons. First, the intermediate states are not pure band states, but also higher
lying exciton states. Second, the assumptions that the deformation potential, which is
used to describe the phonon interaction cannot be taken as a constant, but must be
allowed to depend on the phonon wave vector $\vecs{Q}$. Our theoretical analysis is strongly
substantiated by experimental results, which indeed show not the expected square
root behavior but the absorption coefficient rises more strongly at higher photon
energies. A line shape fit of the absortion then allows to determine precise values for
the deformation potential and its $\vecs{Q}$-dependence. Since the green excitons are coming
from the same valence band states as the yellow excitons, their absorption
processes are closely connected. Therefore, we are able to describe the complete
absorption band of the yellow and green series without additional parameters and
obtain excellent agreement with experiment.

Our results also have practical interest for the use of Cu$_2$O in solar cells, as the
absorption coefficient determines the cell efficiency. In a recent paper \cite{malerba2011absorption}, a
detailed analysis of the whole absorption of Cu$_2$O up to the blue and violet exciton
states was performed, but the authors used the simple square root dependence of
the absorption coefficient and introduced ad hoc values for the deformation potentials
for the green excitons to obtain a fit to the experimental spectrum, making their
analysis invalid.

The paper is organized as follows: In the first section, we discuss the symmetry
properties of Cu$_2$O relevant to the phonon-assisted absorption process. In the
second paragraph the theoretical analysis is presented, while the next section
discusses the experimental procedures. Then we discuss how to obtain the
deformation potential from the fit of the theoretical expressions to the experimental
results. In the last section, we extend the analysis to the green exciton states and
discuss the results.

\section{Symmetries in cuprous oxide}

To comprehend the composition of effects contributing to the excitonic absorption spectra of 
cuprous oxide, we require some basic knowledge of its band structure. 

\begin{figure}[t]
 \includegraphics{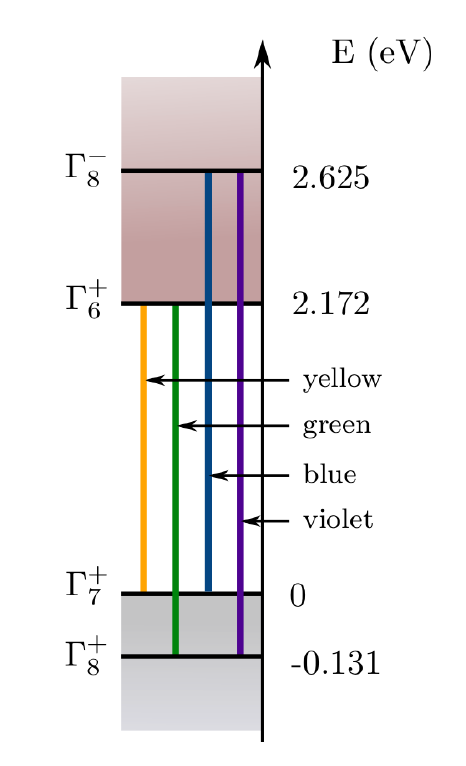}
 \caption{The four exciton series in cuprous oxide with the respective band energies and symmetries at the $\Gamma$ point.}
 \label{fig:01}
\end{figure}

The highest valence band stems from Cu $3d$ orbital with symmetry 
$\Gamma_3^+ \oplus$ $\Gamma_5^+$ at the $\Gamma$ point. Under the crystal field they 
split into the upper $\Gamma_5^+$ and lower $\Gamma_3^+$ bands. The $\Gamma_5^+$ 
bands splits via spin-orbit interaction further into a nondegenerate upper $\Gamma_7^+$ 
band and lower, twofold degenerate  $\Gamma_8^+$ bands with a splitting of 
$\Delta_{so} = 131\,\mathrm{meV}$ at zone center. The lowest conduction band originates 
from the Cu $4s$ orbital, hence possesses a $\Gamma_1^+$ symmetry and becomes a $\Gamma_6^+$ 
band under consideration of spin-orbit interaction. It is followed by a $\Gamma_3^-$ band 
($\Gamma_8^-$ respectively, when including spin), that stems from the Cu $4p$ orbital, which 
is known from band structure calculations~\cite{robertson1983electronic}.
There are also higher located conduction bands with $\Gamma_4^-$ symmetry that are formed by 
the Cu $4p$ orbital.

For our purposes, we are only interested in the two lowest conduction bands 
($\Gamma_6^+ \oplus\Gamma_8^-$) and two highest valence bands 
($\Gamma_7^+\oplus\Gamma_8^+$), since they form the four known exciton series of 
cuprous oxide: the yellow ($\Gamma_6^+\otimes\Gamma_7^+$), green 
($\Gamma_6^+\otimes\Gamma_8^+$), blue ($\Gamma_8^-\otimes\Gamma_7^+$) and violet
($\Gamma_8^-\otimes\Gamma_8^+$). They are visualised in Fig.~\ref{fig:01}. 

Symmetries play an important role as they limit the possibilities for transition between 
the different bands. The symmetry of any respective exciton state is given by
\begin{align}
 \Gamma_\mrm{exc} = \Gamma_\mrm{env}\otimes\Gamma_\mrm{c}\otimes\Gamma_\mrm{v}
\end{align}
To enter any excitonic state directly, the exciton symmetry $\Gamma_\mrm{exc}$
requires to coincide with the symmetry of the respective transition operator. In the $O_h$
group the dipole operator $\vecs{p}$ possesses the symmetry $\Gamma_4^-$, the operator 
$(\vecs{e}\cdot \vecs{p})(\vecs{k}\cdot\vecs{r})$ yields the symmetry 
$\Gamma_3^+\oplus\Gamma_4^+\oplus\Gamma_5^+$, 
which corresponds to the electric quadrupole ($\Gamma_3^+\oplus\Gamma_5^+$) and the magnetic 
dipole transitions ($\Gamma_4^+$). Regarding the excitonic envelope
$\Gamma_\mrm{env}$, states with an S-like character bear $\Gamma_1^+$ symmetry, while
P-like states show $\Gamma_4^-$ symmetry.
The yellow S excitons then split further into  
$\Gamma_1^+ \otimes (\Gamma_6^+ \otimes \Gamma_7^+) = \Gamma_2^+\oplus\Gamma_5^+$ 
via exchange interaction. While the orthoexcitons ($\Gamma_5^+$) are at least quadrupole 
active, the paraexcitons ($\Gamma_2^+$) are inexcitable with light.
The same relation holds true for the green S-excitons, which split as
$\Gamma_1^+ \otimes (\Gamma_6^+ \otimes \Gamma_8^+) = \Gamma_3^+\oplus\Gamma_4^+\oplus\Gamma_5^+$.
On the other hand the dipole transition to the higher located blue and violet S-states
is possible as they decompose to
$\Gamma_1^+ \otimes (\Gamma_8^- \otimes \Gamma_7^+) = 
\Gamma_3^-\oplus\Gamma_4^-\oplus\Gamma_5^-$ for blue and
$\Gamma_1^+ \otimes (\Gamma_8^- \otimes \Gamma_8^+) = 
\Gamma_1^-\oplus\Gamma_2^-\oplus\Gamma_3^-\oplus2\Gamma_4^-\oplus2\Gamma_5^-$
for violet.
For P-type excitons on the other hand, both the yellow
$\Gamma_4^- \otimes (\Gamma_6^+ \otimes \Gamma_7^+) = 
\Gamma_2^-\oplus\Gamma_3^-\oplus\Gamma_4^-\oplus2\Gamma_5^-$
and the green 
$\Gamma_4^- \otimes (\Gamma_6^+ \otimes \Gamma_8^+) = 
\Gamma_1^-\oplus\Gamma_2^-\oplus2\Gamma_3^-\oplus3\Gamma_4^-\oplus3\Gamma_5^-$
series are dipole active.
The beforementioned highly excited states up to 
$n=25$ \cite{kazimierczuk2014giant} are the yellow series P excitons.

As we will see in the upcoming part, most of the absorption background superpositioning 
with the exciton resonances arises from the yellow S excitons. The prevalent contribution 
in the spectra, however, comes from the phonon-assisted absorption process. As mentioned, 
the dipole excitation from either the blue and violet S-excitons is 
possible. Treating the transition into the $\Gamma_8^-$ band (or its respective exciton 
states) as a virtual state with a successive absorption or emission of a phonon, provided 
it has the correct symmetry, we are able to access the yellow (and green) S exciton states. 
Beyond the $\Gamma_8^-$ conduction band, the next closest band that fulfills the necessary 
symmetry and parity restrictions to allow 
for a dipole transition into S-states would be a $\Gamma_4^-$ valence band at around $\sim-5\,\mathrm{eV}$ 
\cite{robertson1983electronic}. As we will see in the next paragraph, the strength of 
absorption contribution depends on the energy difference between the incoming light and 
the virtual state. Approximating the photon energy being around the yellow gap energy the 
ratio between the two dipole allowed states is 
$\left|\frac{\Delta E_{8^-,6^+}}{\Delta E_{4^-,7^+}}\right|^2 \simeq 0.01$, which however, 
might be compensated by a very strong electron-phonon interaction (as will be the case with 
the $\Gamma_4^-$ phonon).

Cuprous oxide features 6 atoms in the primitive unit cell, hence there are 18 phonon branches 
\cite{dawson1973dielectric}. The symmetry of the final exciton state must be contained in 
the direct product of the $\Gamma_4^-$ and the corresponding phonon symmetry. Utilising 
multiplication tables of the $O_h$ group \cite{koster1963properties} it is easy to show that 
a $\Gamma_5^+$ exciton can couple to all odd parity phonons, while a $\Gamma_4^+$ state 
couples to all odd parity phonons exept the $\Gamma_2^-$ mode. Excitons with symmetry 
$\Gamma_2^+$ and $\Gamma_3^+$ only couple to $\Gamma_5^-$ 
and $\Gamma_4^-$, $\Gamma_5^-$ phonons, respectively. 
Additionally, to enable a transition, phonon symmetry must also coincide with the 
predetermined transition symmetries of the bands, in the case of the second lowest conduction 
band $\Gamma_8^- \otimes \Gamma_6^+ = \Gamma_3^- \oplus \Gamma_4^- \oplus \Gamma_5^-$. 
From luminescence spectroscopy we know that the $\Gamma_3^-$ optical phonon with an energy 
of $\hbar \omega_{3-} = 13.6 \,\mathrm{meV}$ at zone center is the dominant phonon branch. 
The contributions of all other phonons should be much weaker, except probably the $\Gamma_4^-$ 
LO phonon at $\hbar \omega_{4-} = 82.1 \,\mathrm{meV}$. Note that the coupling mechanism of 
all phonon modes is via the optical deformation potential, since Fr\"ohlich interaction can only give rise to intraband transitions due to the orthonormality of the Bloch functions
\footnote{The scalar potential of the LO interaction does not modify the Bloch functions, hence can be extracted from the transition matrix elements.}.
The even parity $\Gamma_5^+$ mode can in principle also contribute to the absorption by a 
wave vector dependent deformation potential, which has odd parity \cite{yu1975resonance}.


\section{Theoretical treatment}
Starting with the transition from excitonic vacuum to an exciton state $\mu$ with $\mu$ 
containing the set of quantum numbers $(n,\ell,m)$ of the final state, the transition 
probability can be derived by second order perturbation theory as
\begin{align} \label{eq:001}
  P_\mathrm{0,\mu}(\vecs{k},\omega) =& \frac{2\pi}{\hbar} \sum_{\vecs{Q},\lambda}\left|\sum_\nu  \frac{\langle \Psi_{\mu,\vecs{Q+k}}|h_{\lambda,\vecs{Q}} |\Psi_{\nu,\vecs{k}}\rangle\langle \Psi_{\nu,\vecs{k}} | h_\mathrm{ph} |\Psi_0\rangle}{E_\nu (\vecs{k}) - \hbar \omega}\right|^2 \nonumber \\
  &\times \,\delta \big[E_\mu(\vecs{Q+k}) \mp \hbar\omega_{\lambda,\vecs{Q}} -\hbar\omega \big] \,,
\end{align}
for the absorption or emission of a phonon, respectively. The two transition elements consist 
of the electron-radiation interaction $h_\mathrm{ph}$ and the phonon interaction hamiltonians 
$h_{\lambda,\vecs{Q}}$, with $\lambda,\vecs{Q}$ denoting the associated phonon type and its 
momentum. $E_i(\vecs{k})$ represents the energy dispersion of state $i$, which is usually 
expressed in terms of the effective mass approximation $\frac{\hbar^2 k^2}{2 M_i}$, with $M_i$ 
being the respective excitonic mass. 

We start by assuming the electron-radiation interaction in electric dipole approximation 
$h_\mathrm{ph} = \frac{\mathrm{e}}{m_0}\vecs{A}\cdot\vecs{p}$ since excitations over higher order 
processes (i.e. quadrupole excitation etc.) are negligibly small. As will be seen later, the 
main contribution to the phonon-assisted absorption stems from the 1S excitons of the yellow 
and green series, so we will restrict ourselves to final states with $\ell = 0$. Dipole transitions 
to states of the yellow and green series with S symmetry are forbidden, however this is not the 
case for the subsequent S series' of blue and violet excitons. The blue exciton series can be 
associated with the yellow series, since they share the same $\Gamma_7^+$ valence band, while the 
violet series share the $\Gamma_8^+$ valence bands with the green series. An excitation of 
$\ell=1$ blue/violet excitons is inhibited by the negative parity of the P envelope, and higher 
order angular momenta are considered negligible. Our virtual states therefore consist solely of 
blue/violet S exciton states (depending on the final state being yellow or green respectively).

\begin{figure}[thb]
 \centering
 \includegraphics{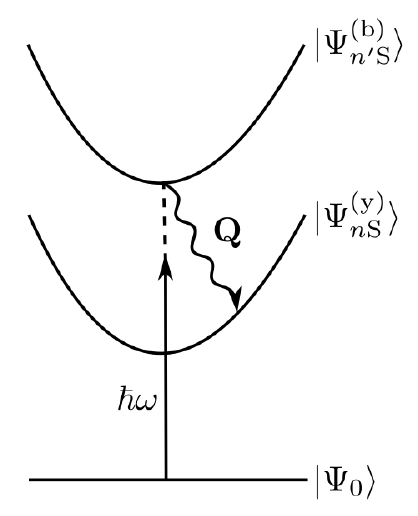}
 \caption{A phenomenological depiction of the phonon-assisted transition as it is characterized by Eq.~(\ref{eq:003}). }
 \label{fig:02}
\end{figure}

The experimental spectra are taken at a crystal temperature of around $2\,\mathrm{K}$, where the 
occupation number of optical phonons converges to zero. Hence, we limit our examination to phonon 
emission. Furthermore, we consider the photon momentum $\vecs{k}$ to be negligibly small. The 
transition probability for the yellow excitons (as seen in Fig.~\ref{fig:02}) then takes the form
\begin{align} \label{eq:002}
 P_{0,n\mrm{S}}^\mrm{(y)}(\omega)  =& \sum_{\lambda} \bar{P}_{n,\mrm{y}}^\lambda(\omega) 
 \end{align}
 \begin{align}\label{eq:003}
  &\bar{P}_{n,\mrm{y}}^\lambda(\omega) = \frac{2\pi \mathrm{e}^2}{\hbar \,m_0^2}\sum_{\vecs{Q}} \nonumber \\
  &\!\times\left|\sum_{n'}  \frac{\langle \Psi_{n\mrm{S},\vecs{Q}}^\mrm{(y)}|h_{\lambda,\vecs{Q}} |\Psi_{n'\mrm{S},0}^\mrm{(b)}\rangle \! \langle \Psi_{n'\mrm{S},0}^\mrm{(b)} | \vecs{A}\cdot \vecs{p} |\Psi_0\rangle}{E_{n'\mrm{S}}^\mrm{(b)} (0) - \hbar \omega}\!\right|^2  \nonumber \\ 
  &\!\times\delta \big[E_{n\mrm{S}}^\mrm{(y)}(\vecs{Q}) + \hbar\omega_{\lambda,\vecs{Q}} -\hbar\omega \big]  \,.
\end{align}
The transition for the green series is equivalent to Eqs.~(\ref{eq:002}), (\ref{eq:003}) by switching yellow to green ($\mrm{y}\rightarrow\mrm{g}$) and blue to violet ($\mrm{b} \rightarrow\mrm{v}$). For the sake of clarity though, we restrict the calculation to the yellow series. The phonon interaction matrix element in Eq.~(\ref{eq:003}) can be rewritten in terms of the Bloch functions $\psi_{n,\vecs{k}}$ of the associated bands as
\begin{align}\label{eq:004}
 \langle \Psi_{n\mrm{S},\vecs{Q}}^\mrm{(y)} & |h_{\lambda,\mathbf{Q}} |\Psi_{n'\mrm{S},0}^\mrm{(b)} \rangle = \nonumber \\ 
 & \sum_{\vecs{q,q'}} \vphi_{n\mrm{S},\vecs{q}}^\mrm{(y)}\,\vphi_{n'\mrm{S},\vecs{q'}}^\mrm{(b)} \, \langle \psi_{6c,\frac{\vecs{Q}}{2}+\vecs{q}}|h_{3,\vecs{Q}} |\psi_{8c,\vecs{q'}}\rangle \, \nonumber \\  
 & \times\langle \psi_{7v,\frac{\vecs{Q}}{2}-\vecs{q}} |\psi_{7v,-\vecs{q'}}\rangle \nonumber\\
 =& \sum_\vecs{q} \vphi_{n\mrm{S},\vecs{q}}^\mrm{(y)}\,\vphi_{n'\mrm{S},\vecs{q}-\frac{\vecs{Q}}{2}}^\mrm{(b)} \, \langle \psi_{6c,\frac{\vecs{Q}}{2}+\vecs{q}}|h_{3,\vecs{Q}} |\psi_{8c,\vecs{q}-\frac{\vecs{Q}}{2}}\rangle\,.
\end{align}
We are expanding the Bloch functions of the matrix elements in Eq.~(\ref{eq:004}) around $\vecs{q}=0$ and express them via a deformation potential $D_{\lambda;ij}$
\begin{align}\label{eq:005}
  \langle \psi_{6c,\frac{\vecs{Q}}{2}}|h_{\lambda,\vecs{Q}} |\psi_{8c,\frac{\vecs{Q}}{2}}\rangle &= D_{\lambda;68} (\vecs{Q}) \sqrt{\frac{\hbar}{2\Omega\,\rho\,\omega_\lambda}} \,,
\end{align}
with $\Omega$ being the crystal volume, and $\rho$ the density of Cu$_2$O.
Consequently, the remainder of the sum reads as $\sum_\vecs{q} \vphi_{n\mrm{S},\vecs{q}}^\mrm{(y)}\,\vphi_{n'\mrm{S},\vecs{q}-\frac{\vecs{Q}}{2}}^\mrm{(b)}$. We assume $\vphi_{n\mrm{S},\vecs{q}}^{(i)}$ to be hydrogen like envelope functions in momentum space. The sum can be evaluated by either simply inserting the momentum hydrogen wave functions \cite{szmytkowski2012alternative}, or treating the expression as a convolution and integrate their product in position space. These convolution functions between different excitonic envelopes are know in the theory of phonon scattering as overlap functions \cite{toyozawa1958theory}. In our case of S type envelopes the spherical harmonics only introduce a factor of $1/(4\pi)$, which in both cases leaves us with a single integral which can be evaluated analytically. For the latter approach we would get
\begin{align}\label{eq:006}
 \mathcal{S}_{n,n'}^\mrm{(y,b)} (Q) =& \sum_\vecs{q} \vphi_{n\mrm{S},\vecs{q}}^\mrm{(y)}\,\vphi_{n'\mrm{S},\vecs{q}-\frac{\vecs{Q}}{2}}^\mrm{(b)} \nonumber \\
		 =&  \frac{2}{Q} \int\limits_0^\infty \mathrm{d}r \; r \,R_{n\mrm{S}}^\mrm{(y)}(r)\, R_{n'\mrm{S}}^\mrm{(b)}(r)\,\sin \frac{Q\,r}{2} \,,
\end{align}
 with $R_{n\mrm{S}}^{(i)}$ being the modified radial hydrogen wave functions. For the dominant transition over 1S states we get
 \begin{align} \label{eq:007}
  \mathcal{S}_{1,1}^\mrm{(y,b)} (Q)= \frac{2^7 \beta^{3/2}(1+\beta)}{\left(4 (1+\beta)^2 + a_\mrm{y}^2 \beta^2\, Q^2 \right)^2} \,,
 \end{align}
where $\beta = a_\mrm{b}/a_\mrm{y}$ and $a_\mrm{y}$, $a_\mrm{b}$ are the excitonic Bohr radii of the yellow and blue series. 

The electron dipole interaction matrix element corresponds to the textbook solution
\begin{align}\label{eq:008}
 \langle \Psi_{n'\mrm{S},0}^\mrm{(b)} | \vecs{A}\cdot \vecs{p} |\Psi_0\rangle = \; A_0\,\vphi_{n'\mrm{S}}^\mrm{(b)}(\vecs{r}\!=\!0) \; p_{78} \,,
\end{align}
with the dipole transition element between Bloch states $p_{78} = \langle u_{8c,\vecs{q}}  | \vecs{e}\cdot \vecs{p}|u_{7v,\vecs{q}}\rangle$, which is considered to not vary significantly over $\vecs{q}$. $\vphi_{n',\mrm{S}}^\mrm{(b)}(\vecs{r})$ is the hydrogen like S envelope function in position space. For $\vecs{r}=0$ it abides to $\vphi_{n',\mrm{S}}^\mrm{(b)}(\vecs{r}\!=\!0) = \big(\pi\,(a_\mrm{b}\,n')^{3}\big)^{-1/2}$.

Inserting the Eq.~(\ref{eq:004}) to (\ref{eq:008}) into Eq.~(\ref{eq:003}) we arrive at
\begin{align}\label{eq:009}
 \bar{P}_{n,\mrm{y}}^\lambda(\omega) =& \frac{\mrm{e}^2 A_0^2}{\Omega m_0 \rho\, \omega_\lambda a_\mrm{b}^3} \frac{\left| p_{78}\right|^2}{m_0} \nonumber  \\
      & \times \sum_\vecs{Q} \left|  D_{\lambda;68} (\vecs{Q})\right|^2 \left|\sum_{n'} \frac{\mathcal{S}_{n,n'}^\mrm{(y,b)} (Q)}{n'^{3/2}\,\big(E_{n'\mrm{S}}^\mrm{(b)}-\hbar \omega\big)}\right|^2 \nonumber \\
      & \times \delta \big[E_{n\mrm{S}}^\mrm{(y)}(\vecs{Q}) + \hbar\omega_{\lambda,\vecs{Q}} -\hbar\omega \big]\,.
\end{align}
The absorption coefficient is defined as
\begin{align}
 \alpha_{n,\mrm{y}}^{\lambda}(\omega) =& \frac{2\hbar}{\varepsilon_0 n_R c \omega A_0^2} \bar{P}_{n,\mrm{y}}^\lambda(\omega) \label{eq:011} \,,
\end{align}
with $n_R$ being the refractive index around the excitation energy. 
To simplify the calculation, we assume that both the deformation potential and the 1S exciton dispersion have spherical symmetry, any deviation can in principle be treated, but would make the following integration more complex. In addition, the phonons of interest are optical phonons with only marginally varying energy dispersions \cite{yu1975resonance}, hence their energies $\omega_{\lambda}$ will be considered constant. Under these circumstances the sum in Eq.~(\ref{eq:009}) can be evaluated. Thus we obtain
\begin{align}\label{eq:012}
 \alpha_{n,\mrm{y}}^{\lambda}(\omega) =&\frac{ \mrm{e}^2}{\pi^2 \hbar\, \rho\, \varepsilon_0 n_R c\, a_\mrm{b}^3 } \frac{M_\mrm{y}}{m_0}\,\frac{q_n^\lambda}{\omega\,\omega_\lambda} \nonumber \\
 &\times \frac{|p_{78}|^2}{m_0} \left|D_{\lambda;68} (q_n^\lambda) \right|^2 \nonumber \\
 & \times \left| \sum_{n'}\frac{ \mathcal{S}_{n,n'}^\mrm{(y,b)} (q_n^\lambda) \; }{ n'^{3/2} \;\big(E_{n'\mrm{S}}^\mrm{(b)}-\hbar \omega\big)}\right|^2 \; \Theta(q_n^\lambda) \,,
\end{align}
with
\begin{align}\label{eq:013}
 q_n^\lambda (\omega) = \sqrt{\frac{2 M_\mrm{y}}{\hbar^2}} \sqrt{\hbar\omega -\hbar\omega_\lambda -E_{n\mrm{S}}^\mrm{(y)}} \,.
\end{align}
The square root behaviour of the textbook solution is still recognisable and is embedded in $q_n^\lambda (\omega)$, though Eq.~(\ref{eq:012}) shows additional photon energy dependencies, such as the convolution of wave functions and the momentum dependent deformation potential. Furthermore, it is not depending on the Bohr radius of the final state, the yellow exciton, but on that of the blue exciton. Therefore, the dependence on the quantum number $n$ of the yellow states is solely in the overlap factors $\mathcal{S}_{n,n'}^\mrm{(y,b)}$.

\section{Experimental data}
Two samples were prepared to obtain absorption spectra in different photon energy ranges\footnote{Ref. \cite{naka2005thin}, copyright 2005 The Japan Society of Applied Physics.}. For the observation of yellow exciton series, a thin slab of natural crystal Cu$_2$O was cut, mechanically polished, and then surface treated by NH$_4$OH. The thickness was measured as 160 $\mu$m by a caliper. For the observation of green exciton series, a thinner sample was grown by the melt-growth method as described in \cite{naka2005thin}. Cu$_2$O powder was sandwiched between two MgO plates of 500 $\mu$m thickness, and then heated up to 1523 K above the melting point of Cu$_2$O. A wedge-shaped Cu$_2$O film was formed between the substrates after cooling down to room temperature. The film thickness was measured by a stylus profiler after removal of the top substrate. At the point of measurement film thickness is $10\,\mrm{\mu m}$.

Absorption spectra were taken with samples at 2 K, immersed in superfluid helium in a cryostat. The white light from a halogen lamp, transmitted through the sample, was measured by a Peltier-cooled CCD camera (Wright Instruments) equipped at the back of a 25 cm monochromator with a 1200 g/mm grating blazed at 500 nm (JASCO CT-25T).  Estimation of the reflectivity of light at the sample and substrate surfaces was difficult. In calculating the absorption coefficients
\begin{align}\label{eq:013b}
 \alpha=-\ln\left[\frac{I_t}{b\,I_0}\right]/d \,,
\end{align}
we adjusted the magnitudes of the reference light (by the factor of $b$) so that the transmission at 620 nm wavelength becomes unity.\footnote{The change in refractive index from $2.95$ at the band edge to about $3.15$ \cite{ito1998optical} at the green excitons would give a correction of $2\%$, which is of order of the experimental error.}
Here, $I_t$ and $I_0$ represent light intensities with and without a sample in the optical path, $d$ is sample thickness.

\section{Deformation potential and yellow 1S phonon transition}
Taking a look back at Eq.~(\ref{eq:012}), while providing us with an analytical solution for the absorption of the phonon background, it still harbors two uncertainties: 
The strength and momentum dependency of the deformation potential $D_{\lambda;68}$.
Fortunately the phonon-assisted absorption into the yellow 1S exciton via the $\Gamma_3^-$ LO phonon features a wide and distinctive spectral shape. We will utilize this property to do a fit of Eq.~(\ref{eq:012}) unto the spectral data. In consideration of which blue S states actually contribute to this absorption line, one can easily calculate the ratios:
\begin{align}
 \label{eq:014} \frac{\mathcal{S}_{1,2}^\mrm{(y,b)}}{2^{3/2}\,\mathcal{S}_{1,1}^\mrm{(y,b)}} \lesssim&\; 12\% \,, \\
 \label{eq:015} \frac{\mathcal{S}_{1,3}^\mrm{(y,b)}}{3^{3/2}\,\mathcal{S}_{1,1}^\mrm{(y,b)}} \lesssim&\;  3.7\% \,.
\end{align}
Since already the blue 3S contribution is small, the inner sum in Eq.~(\ref{eq:012}) over the intermediate states is run up to $n'=3$, 
which should be sufficient to safely neglect the contribution from higher blue states.
The deformation potential can be expanded with respect to the square of phonon momentum $Q$ to
\begin{align}
 \label{eq:016} D_{\lambda,68}(Q) =& \; D_{\lambda,68}^{(0)} \;+\; D_{\lambda,68}^{(2)}\,Q^2\; + \ldots \nonumber \\
    \simeq& \; D_{\lambda,68}^{(0)}\left(1 \;+\; \bar{D}_{\lambda,68}^{(2)}\,Q^2\right) \,.
\end{align}
While usually the deformation potential is assumed to be constant, we will show that in case of the $\Gamma_3^-$ phonon this is not the case.
In this paper we will consider the zeroth and first order of Eq.~(\ref{eq:016}). Higher order terms only increase the number of variables to fit and do not improve the result noticeably. For comparison we also fit the standard approach derived by Elliott \cite{elliott1957intensity} of the form
\begin{align} \label{eq:017}
 \alpha_E^\lambda (\omega) \propto \sum_{n=1} \frac{1}{n^3} \sqrt{\hbar\omega -\hbar\omega_\lambda -E_{n\mrm{S}}^\mrm{(y)}} \,.
\end{align}
\begin{figure}
 \centering
 \footnotesize
 \includegraphics{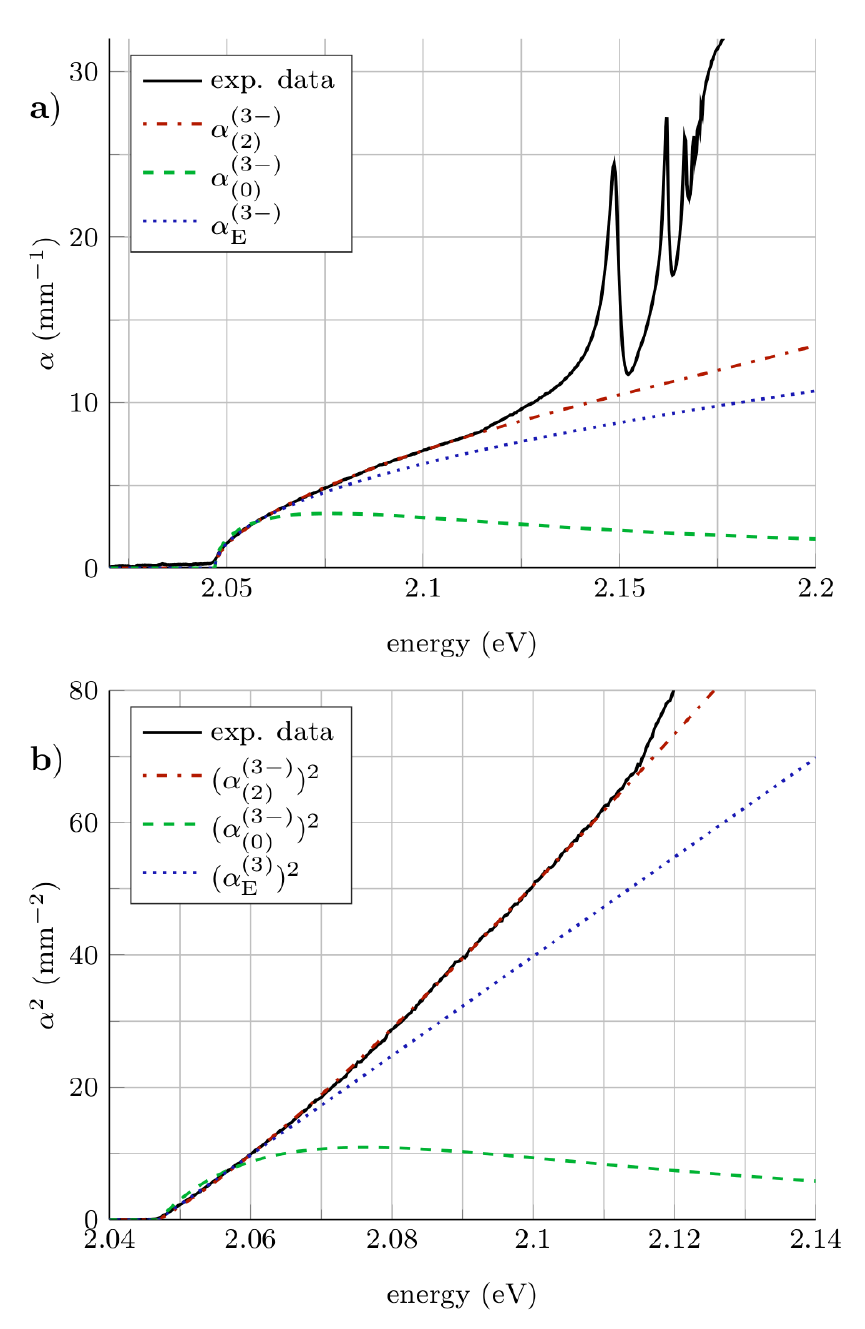}
 \caption{Comparison of different fits for the absorption edge of the phonon-assisted 1S exciton for the sample of $160\,\mrm{\mu m}$ thickness at $2\,\mrm{K}$. While our theory flops for a constant deformation potential, the assertion of a momentum dependent deformation potential fits experimental data very well. In $\vecs{b)}$ the absorption coefficient is squared for better visibility, clearly showing the non-square root behavior of the absorption coefficient for energies $>2.08\, \mrm{eV}$.}
 \label{fig:03}
\end{figure}
The result can be seen in Fig.~\ref{fig:03}. While the assumption of a constant deformation potential $\alpha_{(0)}^{(3-)}$ fails to represent the experimental curve completely, the approach of Elliott $\alpha_{E}^{(3-)}$ fits well for a short energy range close to absorption edge, however deviates with increasing energy. The fit with the momentum dependent deformation potential $\alpha_{(2)}^{(3-)}$ reproduces the spectra neatly up to the start of the overlaying $\Gamma_4^-$ phonon absorption edge, which can be seen as a shoulder around $2.116\,\mrm{eV}$ at the upper right corner of Fig.~\ref{fig:03} \vecs{b)}. 
\\ 
The $\Gamma_4^-$ phonon transition requires either $\Gamma_4^-$ conduction or $\Gamma_2^-\oplus\Gamma_3^-\oplus\Gamma_4^-\oplus\Gamma_5^-$ valence bands\footnote{The consideration of the ``spinless'' symmetries is sufficient here. The spin-including symmetries open up additional transitory channels, however those require a change in the spin configuration, which cannot be inflicted by phonons.}.
While bands with these symmetries exist \cite{kleinman1980self}, we have to keep in mind, that they are located energetically quite far away from the yellow band gap $E_\mrm{g}$ and we are not cognisant of any excitonic properties. Thus, the approach of Eq.~(\ref{eq:012}) is not really suited for their treatment, and since the $\Gamma_4^-$ phonon contribution is small enough, we purposely approximate it in the fashion of Eq.~(\ref{eq:017}). The result of the fit is given in appendix~\ref{minorfits}.
\\We stress, that all other phonons contribute with negligible strength to the yellow absorption band. The inclusion of the $\Gamma_4^-$ phonon transition allows us to describe the phonon-assisted absorption into the 1S yellow exciton very accurately up to the P transitions.
\begin{table}
\centering
\begin{tabular}{lr}
\textbf{Parameter} & \textbf{Value} \\ \hline \hline
density \cite{korzhavyi2011literature} & $\rho = 6.14 \times 10^3 \,\mrm{kg/m^3}$\\
refractive index \cite{brandt2007ultranarrow} & $n_R = 2.94$\\
1S exciton mass (y) \cite{brandt2007ultranarrow} & $M^\mrm{(y)}_\mrm{1S} = 2.61 \,m_0$\\
Bohr radius (y)\cite{schweiner2017even}  & $a_\mrm{y} = 0.81 \,\mrm{nm}$\\
Bohr radius (b)  & $a_\mrm{b} = 1.72 \,\mrm{nm}$\\
1S exciton energy (y) \cite{uihlein1981investigation} \hspace{.5cm} & $E_\mrm{1S}^\mrm{(y)} = 2.033\,\mrm{eV}$\\
1S exciton energy (b) \cite{daunois1966etude} & $E_\mrm{1S}^\mrm{(b)} = 2.569\,\mrm{eV}$\\
$\Gamma_3^-$ phonon energy \cite{yu1975resonance} & $\hbar \omega_{3-} = 13.6\,\mrm{meV}$\\
$\Gamma_4^-$ phonon energy \cite{yu1978resonance} & $\hbar\omega_{4-} = 82.1\,\mrm{meV}$ \\
dipole transition element & $|p_{78}|^2/m_0 = 2.66 \,\mrm{eV}$
\end{tabular}
\caption{Necessary parameters to evaluate Eq.~(\ref{eq:012}) for the transition into the yellow 1S state via the $\Gamma_3^-$ phonon.}
\label{t:01}
\end{table} 

The used parameters can be found in table~\ref{t:01}. The Bohr radius of the blue excitons is not explicitly known, thus for a systematic treatment they were calculated from the binding energies via $a_\mrm{b} = \mrm{Ry}\; a_\mrm{B} /(\mrm{Ry}_\mrm{b} \varepsilon_0)$, with $\mrm{Ry}$ and $a_\mrm{B}$ being the (hydrogen) Rydberg energy and Bohr radius, and $\varepsilon_0 = 7.5$ \cite{kavoulakis1996auger}.
The dipole transition element can obtained from experiments as well as $\vecs{k}\cdot\vecs{p}$ theory (see appendix~\ref{dipolevalue}).
The resulting fit parameters for the deformation potential of the $\Gamma_3^-$ phonon-assisted absorption are
\begin{align}\label{eq:018}
 D_{3;68}^{(0)} &= 25.45 \,\mrm{\frac{eV}{nm}} \,, \\
 \bar{D}_{3;68}^{(2)} &= 0.168 \,\mrm{nm^2}\label{eq:019} \,.
\end{align}
The value for the static deformation potential $D_{3;68}^{(0)}$ is in well accordance to previous estimations \cite{kavoulakis1996auger}.

\section{The spectrum beyond yellow 1S}

\subsection{Extrapolating the previous result}
The practical aspect of the fitted parameters (\ref{eq:018}) and (\ref{eq:019}) is 
the fact that they are applicable for all $\Gamma_3^-$ (and $\Gamma_4^-$) phonon-assisted 
transitions into the yellow S series, i.e. we additionally receive the absorption 
strengths for all $n\geq 2$ states. 
However, the contributions from states with $n>2$ and into the yellow continuum, which 
would start at $2.186\,\mrm{eV}$, is negligible and will not be taken into account.

Another and perhaps more interesting proposition stems from the relation between transition 
elements of the yellow and green series. From group theoretical symmetry considerations 
it can be shown that
\begin{align}\label{eq:020}
 \frac{\alpha_\mrm{g}^{\Gamma_3^-}}{\alpha_\mrm{y}^{\Gamma_3^-}} \; = \;\, \frac{2}{1} \,.
\end{align}
The derivation is shown in appendix~\ref{transitionstrgth}. Therefore we also possess all 
necessary information to describe the $\Gamma_3^-$ phonon-assisted transitions into the 
green S series. At zone center, symmetry considerations predict that the green series is 
supposed to consist of three distinct states $\Gamma_3^+$, $\Gamma_4^+$ and $\Gamma_5^+$. 
However, only the $\Gamma_5^+$ orthoexciton is accessible.

Beyond that, we consider the absorption into the yellow P exciton states phenomenologically 
to achieve a well-rounded depiction of the absorption spectrum. The excitonic resonances 
are accessed via a forbidden dipole transition\cite{elliott1957intensity}, with the oscillator 
strength varying with principal quantum number as $(n^2-1)/n^5$. The lineshape of the P 
excitons can be described via asymmetric Lorentzians as derived by Toyozawa \cite{toyozawa1958theory} 
and the successive transition into the yellow continuum is given by the Sommerfeld enhanced 
direct forbidden absorption \cite{kuper1963polarons}. 
Recently, we have shown  that the continuum absorption is shifted by an energy $\Delta_\mrm{c}$ into 
the P states due to plasma screening of charged residual impurities. In addition, the continuum absorption develops an Urbach tail behavior 
$\exp((\hbar\omega-E_\mrm{g})/E_\mrm{U})$ with $E_U$ being the Urbach parameter \cite{heckotter2017plasma}. 
 These contributions are conglomerated into 
$\alpha_\mrm{P}$, for details, see appendix~\ref{minorfits}.

\begin{figure}
 \centering
 \includegraphics{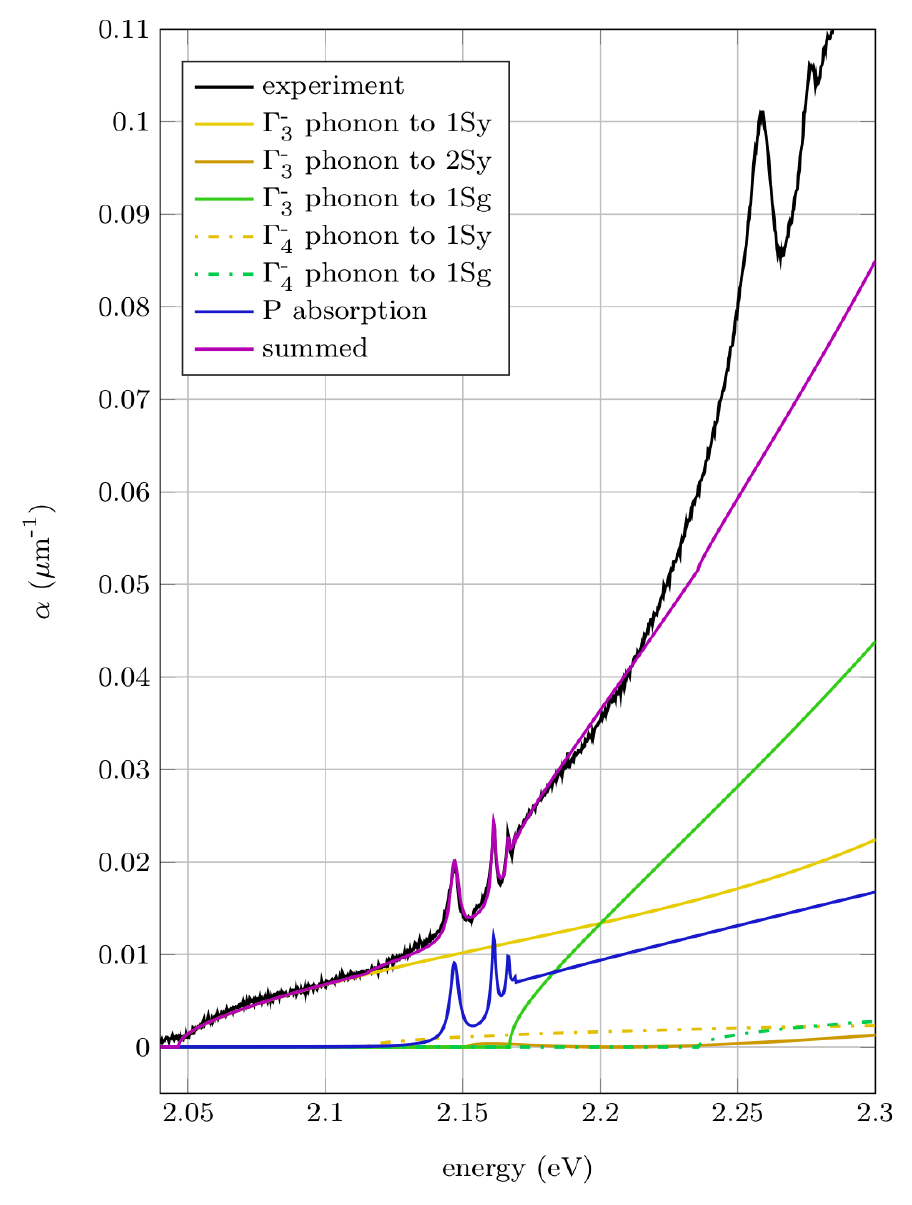}
 \caption{The phonon-assisted absorption edge of the significantly contributing absorption participants $\alpha_\mrm{tot}$ 
 . The sum of the individual participations is compared with the experimental spectrum of sample with thickness 10$\,\mrm{\mu m}$ at $2\,\mrm{K}$. }
 \label{fig:05}
\end{figure}

With the the fitted parameters in (\ref{eq:018}) and (\ref{eq:019}) we will now attempt to 
depict the absorption spectrum beyond the yellow band gap $E_\mrm{g}$. The total absorption 
coefficient only requires only the absorption channels that significantly participate and is 
therefore composed of
\begin{align}\label{eq:021}
 \alpha_\mrm{tot} = \alpha_{1,\mrm{y}}^{\Gamma_3^-} +\alpha_{2,\mrm{y}}^{\Gamma_3^-}+\alpha_{1,\mrm{g}}^{\Gamma_3^-}+\alpha_{1,\mrm{y}}^{\Gamma_4^-}+\alpha_{1,\mrm{g}}^{\Gamma_4^-}+\alpha_\mrm{P}\,.
\end{align}
The remaining parameters needed to evaluate $\alpha_\mrm{tot}$ are listed in table~\ref{t:02}. The green and violet Bohr radii are obtained in the same fashion, as it was done for the blue exciton. The yellow 2S exciton mass stems from the sum of effective electron and hole mass. Due to the almost equal binding energy of the yellow 1S paraexciton and the green 1S orthoexciton, and since they experience the same screening, it is expected that both masses are about equal. The result is shown in Fig.~\ref{fig:05}.
\begin{table}
\begin{center}
\begin{tabular}{l r}
\textbf{Parameter} & \textbf{Value}\\ \hline \hline
2S exciton mass (y)  & $M^\mrm{(y)}_\mrm{2S}  =1.56 \,m_0$\\
1S exciton mass (g)  & $M^\mrm{(g)}_\mrm{1S}  = 2.61 \,m_0$\\
Bohr radius (g)  & $a_\mrm{g} = 0.74\,\mrm{nm}$\\
Bohr radius (v)  & $a_\mrm{v} = 1.36\,\mrm{nm}$\\
2S exciton energy (y) \cite{uihlein1981investigation} & $E^\mrm{(y)}_\mrm{2S} = 2.138\,\mrm{eV}$\\
1S exciton energy (g)  \cite{uihlein1981investigation} \hspace{1cm} & $E^{\mrm{(g)}}_\mrm{1S} = 2.154\,\mrm{eV}$\\
1S exciton energy (v) \cite{daunois1966etude} & $E^\mrm{(v)}_\mrm{1S} = 2.715\,\mrm{eV}$ 
\end{tabular}
\end{center}
\caption{Parameters to evaluate Eq.~(\ref{eq:021}) for the remaining transitions, shown in Fig.~\ref{fig:05}. }
\label{t:02}
\end{table} 

We recognise that the absorption spectrum is mainly constructed out of the $\Gamma_3^-$ phonon-assisted
transition of the yellow and green 1S exciton state, as well as the excitonic resonances and the 
absorption into the continuum at the band edge. While at an energy range around $2.2\,\mrm{eV}$ 
the summed up absorption coefficient $\alpha_\mrm{tot}$ appears to slightly overestimate the 
absorption, beyond $2.22\,\mrm{eV}$ the experimentally measured spectrum shows a steady
increase in the absorption slope that is not reproduced in our theory.

\subsection{Discussion}
\noindent We will first discuss the overshot of the theoretical results 
over the experimental data at $2.2\,\mrm{eV}$.

The biggest issue in extrapolating the result of the yellow 1S exciton to the 
green 1S state is the fact that two of the essential exciton parameters are not
explicitly known. The excitonic Bohr radii used throughout this work are all
extracted from the respective binding energies.
They come into play via the overlap functions Eq.~(\ref{eq:007}).
The other parameters that determine the strength of the absorption
are the exciton translational masses according to Eq.~(\ref{eq:013}). 
Note that a change of 10\% in the masses would alter the absorption by 15\%.
While for the 1S yellow
state the mass has been determined experimentally to be $M_\mrm{1Sy}=2.61 \,m_0$ 
\cite{brandt2007ultranarrow}, for the 2S yellow and the green exciton 
series it is not known. However, for excitons with large principal quantum
numbers, which are composed of valence and conduction band states near the zone
center, the approximation $M_X =m_c + m_v$ holds true, where $m_c$ and $m_v$ are the effective
masses at $k=0$. Considering the effective mass of the conduction band $m_6 = 0.985\,m_0$,
and that of the $\Gamma_7^+$ hole of $m_{7}=0.575$, both from time-resolved cyclotron
resonance \cite{naka2012time}, we obtain for the 2S yellow exciton a mass of $1.56\,m_0$.

For the green 1S state, whose wave function extends due to its large binding energy far into the Brillouin zone, the influence of the non-parabolicity of the $\Gamma_8^+$ valence bands on Bohr radius and effective mass are expected to be similar to that for the yellow 1S exciton. Using the heuristical relation between binding energy and translational mass as for the yellow state, we obtain a mass of $M_{1Sg}=2.61\,m_0$. The Bohr radius is given in table~\ref{t:02}. Of course, all these quantities are only first order approximations, and require extensive future work to improve. Theoretically, the Bohr radii and translational masses can be obtained from the solution of the $\vecs{K}$-dependent effective mass equations, where $\vecs{K}$ is the center-of-mass wave vector, which will be the topic of a forthcoming paper. Experimentally, the masses can be obtained from resonance Raman studies involving the green P and the green 1S states, similar to that reported by Yu \textit{et al.} in the 1970's \cite{yu1978resonance} for the yellow exciton states. Here especially Raman processes involving acoustical phonons are of interest, since
their Raman shift depends on their momentum, thus giving directly the dispersion of the green 1S states. These experiments would also clarify a possible contribution of a $\Gamma_5^+$ phonon to the absorption. 


The most interesting finding of Fig.~\ref{fig:05} is the steady increase of absorption in the region 
around $2.22\,\mrm{eV}$. This cannot be explained by simple modifications within this framework, i.e. 
the choice of parameters or of the wave function used.
Theoretically, it could be explained by adding additional absorption channels, e.g. by introducing 
additional phonon interactions or exciton resonances, but it can easily be shown that this is not
the case here. The only phonon resonance with a suitable energy to compensate the missing absorption 
would be the $\Gamma_5^+$ phonon at $63.8\,\mrm{meV}$. However, the possibility for a scattering into 
the green states involving the $\Gamma_5^+$ phonon can be ruled out, since the phonon possesses the 
wrong parity and such a process must also occur in comparable strength for the yellow 1S state but 
could not be identified in the analysis. The existence of a potential second green exciton series,
that could be associated with the $\Gamma_8^+$ light hole band dispersion, was considered, but
since the $\Gamma_8^+$ valence bands are heavily coupled \cite{baldereschi1971energy} no such
additional exciton series should exist. 
Both ideas also seem implausible, since an additional phonon-assisted absorption edge would rise up 
abruptly in the spectrum, while the increase of the slope appears to be continuous.
\\ Currently, the most logical explanation is a dependence of the excitonic parameters on exciton
momentum $\vecs{K}$. 
If the exciton mass is expected to steadily increase with exciton momentum, a smooth increase in 
phonon-assisted absorption with photon energy, as it is seen in the experiment, should be observed 
(cf. Eq.~(\ref{eq:013})). The same happens, if the green exciton Bohr radius increases, as the overlap 
functions (cf. Eq.~\ref{eq:007})  would also increase. A solution of this problem might come from the 
aforementioned study of the $\vecs{K}$ dependent effective mass equations. 

Finally, we will discuss the topic of mixing between yellow and green exciton states, which was fo<und in a recent study of the even excitons in Cu$_2$O \cite{schweiner2017even}. According to this work, the lowest orthoexciton resonance (our 1S yellow state) is a mixture of yellow and green states with 7.2\% green contribution. Taking this into account in our analysis would require the redetermination of the Deformation potential $D_{3;68}$ by refitting the $\Gamma_3^-$ absorption band in the low energy part of the spectrum (cf. Fig.~\ref{fig:03} b), but we expect only a small correction of the order of some percent due to the low green admixture.
Much more pronounced should be the influence of mixing onto the absorption of the yellow ``2s'' and the green ``1S'' state. The former has more than 10\% of green contributions, which would enhance its absorption strength considerably. In contrast, the green 1S state should have only a contribution of green states of about 40\%, which, taking literally, should result in only half the absorption strength (cf. solid green line in Fig.~\ref{fig:05}). Both effects are clearly not consistent with the experimental data. However, a rigorous analysis would require the full wavefunctions of these exciton states, which is an interesting task for future work.

\section{Conclusion}

In an effort to determine the phonon-assisted absorption background around the band edge of Cu$_2$O 
we evaluated the established second order perturbation treatment of Elliott. However, the resulting 
square root behaviour of this textbook solution is not sufficient to reliably reproduce experimentally 
measured spectra for energies much higher than the absorption edge. We reassessed the approach and 
removed three distinct approximations that are not necessarily justifiable. In a semiconductor with 
strong excitonic features, like Cu$_2$O, instead of treating intermediate states as pure band states 
we need to consider the corresponding exciton eigenstates for the virtual transition. Additionally, 
the momentum dependence of the optical phonons deformation potential was taken into account as well 
as the excitation energy dependent denominator. The resulting improved expression of the absorption 
coefficient [Eq.~(\ref{eq:012})] is able to effectively model the $\Gamma_3^-$ phonon-assisted transition 
into the yellow 1S exciton state, the strongest and most distinct phonon-assisted transition. Beyond 
that, we modelled the $\Gamma_4^-$ phonon transition into the yellow 1S state and extrapolated our 
results from the yellow to the green series excitons. This yields a profound description of the 
phonon-assisted absorption background up to the yellow band gap. Beyond that, a sudden increase in 
the experimental absorption spectra is found, that could not be explained with our current theoretical
treatment. The possibility of momentum dependent exciton parameters is discussed as a potential 
origin.

\acknowledgments We gratefully acknowledge 
support by the Collaborative Research Centre SFB 652/3 'Strong correlations in 
the radiation field' funded by the Deutsche
Forschungsgemeinschaft.

\appendix

\section{Fitting results for the minor absorption contributions}\label{minorfits}

\textit{The $\Gamma_4^-$ phonon transition:}
\\As previously mentioned, the $\Gamma_4^-$ phonon scattering is fitted with the square root solution of Eq.~(\ref{eq:017}), as it couples to a multitude of higher (lower) located conduction (valence) bands, of which we cannot distinguish the individual transitions. Since its absolute contribution to the spectrum is marginal, we are content with only considering the $n=1$ state. Utilising Eq.~(\ref{eq:013}) we get
\begin{align}
 \alpha_E^{\Gamma_4^-}(\omega) = C_4\;\, q_1^{\Gamma_4^-}\!(\omega) \,,
\end{align}
with the corresponding fit parameter
\begin{align}
 C_4 = 6.56 \times 10^{-7} \,.
\end{align}

\textit{The yellow P-absorption:}
\\ The P-absorption is divided into three separate parts
\begin{align}
 \alpha_\mrm{P} = \alpha_\mrm{Pcont} + \alpha_\mrm{Urbach} + \sum_{n=2}^4 \alpha_{n\mrm{P}} \,.
\end{align}
The continuum is given by\cite{kuper1963polarons}
\begin{align}
 \alpha_\mrm{Pcont}(\omega) &= C_\mrm{yP} \frac{(\hbar\omega-\tilde{E}_\mrm{g})^{3/2}}{\hbar\omega} \; \frac{\gamma \,e^\gamma}{\sinh \gamma} \left(1+\frac{\gamma^2}{\pi^2}\right) \,,
\end{align}
with
\begin{align}
 \gamma = \sqrt{\frac{\pi^2 \,\mrm{Ry_y}}{\hbar\omega-\tilde{E}_\mrm{g}}}\,,
\end{align}
and the yellow Rydberg energy\cite{schone2016deviations} $\mrm{Ry_y}= 87\,\mrm{meV}$. 
The renormalized band gap $\tilde{E}_\mrm{g} = E_\mrm{g}+\Delta_\mrm{c}$ reflects the band gap shift due to plasma screening. Roughly,  
$\Delta_\mrm{c}$ can be estimated from the energy of the highest visible P exciton line ($n_\mrm{max}=4$) as 
$-87\,\mrm{meV}/n_\mrm{max}^2$. It depends on the 
sample properties and thus is different for the thick and thin sample. 
The fit yields
\begin{align}
 C_\mrm{yP} = 9.82 \times 10^{-02}\, (\sqrt{\mrm{eV}}\,\mrm{\mu m})^{-1} \,.
\end{align}
The Urbach tail is given by
\begin{align}
 \alpha_\mrm{Urbach}(\omega) = C_\mrm{U} \exp\left(\frac{\hbar\omega-\tilde{E}_\mrm{g}}{E_\mrm{U}}\right) \,\theta(\tilde{E}_\mrm{g}-\hbar\omega) \,,
\end{align}
with $C_\mrm{U} = 7.34 \times 10^{-03} \,\mrm{\mu m}^{-1}$ and $E_\mrm{U} = 9.8 \,\mrm{meV}$. The exciton resonances are described by asymmetric Lorentzians\cite{toyozawa1958theory}
\begin{align}
 \alpha_{n\mrm{P}}(\omega) = C_{n\mrm{P}} \frac{\Gamma_{n\mrm{P}}/2 + 2\xi_n\hbar(\omega-\omega_n) }{(\Gamma_{n\mrm{P}}/2)^2 + \hbar^2(\omega-\omega_n)^2} \,.
\end{align}
The values used are:
\begin{center}
\begin{tabular}{l|ccc}
n & 2P & 3P & 4P\\ \hline\hline
$\hbar\omega_n$\cite{kazimierczuk2014giant}\,$(\mrm{eV})$ & 2.1472 & 2.1612 & 2.16604\\
$C_{n\mrm{P}}\,(10^{-5}\mrm{eV/\mu m})$\hspace{.4cm} & 1.587 & 0.793 & 0.2645\\
$\Gamma_{n\mrm{P}}\,(\mrm{meV})$ & 3.86 & 1.93 & 1.29\\
$\xi_{n\mrm{P}}\,(10^{-3})$ & -4.32 & -4.32 & -4.32
\end{tabular}
\end{center}

\section{Dipole transition element $p_{78}$}\label{dipolevalue}

The dipole transition element of the blue exciton is related to the oscillator strength by
\begin{align}
 \frac{f_\mrm{b}}{\Omega_\mrm{uc}} = \frac{2}{\hbar\omega} \frac{|p_{78}|^2}{m_0} \,\left| \vphi_\mrm{1S}^\mrm{(b)}(0)\right|^2 = \frac{2}{\pi a_\mrm{b}^3\,\hbar\omega} \frac{|p_{78}|^2}{m_0} \,,
\end{align}
with $\Omega_\mrm{uc}=a_\mrm{L}^3$ being the volume of the uni cell, $a_\mrm{L}=0.45\,\mrm{nm}$ is the lattice constant, and $a_\mrm{b}$ is given in table~\ref{t:01}. In \cite{schmutzler2013signatures} the oscillator strength was determined to be $f_\mrm{b} = 1.2\times 10^{-2}$. This yields for the dipole transition element
\begin{align}
 \frac{|p_{78}|^2}{m_0} = 2.726 \,\mrm{eV} \,;
\end{align}
however this approach is reliant on the blue excitons Bohr radius, which is not well known. Therefore, we employ a second derivation to double check the result.
The dipole transition element for the blue exciton series is related to the transition matrix element of the $\Gamma_5^+$ valence and $\Gamma_3^-$ conduction band basis states by \cite{koster1963properties}
\begin{align}\label{eq:025}
 p_{78} = -\sqrt{\frac{2}{3}} \,\langle \varepsilon_3^+ |\,\vecs{p}\,|\gamma_2^- \rangle \,.
\end{align}
The transition matrix element of Eq.~(\ref{eq:025}) appears in the Suzuki-Hensel Hamiltonian \cite{suzuki1974quantum} in the coefficient
\begin{align}\label{eq:026}
 G   &= \frac{2}{m_0} \sum_{\ell = \Gamma_3^-} \frac{|\langle \varepsilon_3 |p_z | \gamma_2^-,\ell \rangle|^2}{E_{5v} -E_\ell}  \,,
\end{align}
coupling all $\Gamma_3^-$ bands to the respective $\Gamma_5^+$ band. The coupling coefficients are directly connected to the three dimensionless parameters $A_i$ ($i=1,2,3$) of the Hamiltonian. Albeit the dipole coupling to a $\Gamma_5^+$ band is possible via four different symmetries $\Gamma_4^-\otimes\Gamma_5^+ = \Gamma_2^-\oplus\Gamma_3^-\oplus\Gamma_4^-\oplus\Gamma_5^-$, and thus there should exist four separate coupling coefficients, band structure calculations of Cu$_2$O show \cite{kleinman1980self}, that no $\Gamma_2^-$ band is located in the near vicinity of the $\Gamma_5^+$ valence band. Therefore when the coupling coefficient for $\Gamma_2^-$ is neglected, the system of equations is solvable. The dimensionless parameters $A_i$ are known from band structure fits \cite{schone2016deviations} and result in a value of 
\begin{align}\label{eq:027}
 G = -2.973 \,.
\end{align}
As there is also only one $\Gamma_3^-$ band in the vicinity of the $\Gamma_5^+$ band, the coupling coefficient $G$ can be associated with the matrix element in Eq.~(\ref{eq:025}), hence
\begin{align}\label{eq:027b}
 \frac{|p_{78}|^2}{m_0} = 2.662 \,\mrm{eV} \,.
\end{align}
Both results are in fairly good agreement. For the estimation of the static deformation potential, we use the result of Eq.~(\ref{eq:027b}).

\section{Phonon-assisted transition strength of the yellow and green series}\label{transitionstrgth}

We are denoting the band to band dipole transition matrix element between $\Gamma_\mrm{5v}^+$ and $\Gamma_\mrm{3c}^-$ bands of Eq.~(\ref{eq:025}) as
\begin{align}\label{eq:app028}
 p_{35} = \langle \varepsilon_3^+ |\,\vecs{p}\, |\gamma_2^- \rangle \,.
\end{align}
We now calculate the relative strength of the dipole transition strength of the blue and violet transition, respectively. We are only interested in the $\Gamma_4^-$ states, as they are the only ones accessible via the dipole operator $\vecs{p}$. The composition of these states is known from the coupling coefficients of the $O_h$ group \cite{koster1963properties}. The resulting dipole transition matrix elements read as
\begin{align}
 \langle 0 |\, \vecs{p}\, | Z\rangle_\mrm{b}\; &= \, -\sqrt{\frac{2}{3}} \;p_{35} \,,\label{eq:app033} \\
 \langle 0 |\, \vecs{p}\, | Z\rangle_\mrm{v,1} &= \, -\sqrt{\frac{6}{5}}\; p_{35} \,,\label{eq:app034} \\
 \langle 0 |\, \vecs{p}\, | Z\rangle_\mrm{v,2} &= \, -\sqrt{\frac{2}{15}}\; p_{35} \,,\label{eq:app035}
\end{align}
For the phonon-assisted transition, we additionally need to consider the transition strength of the phonon process. 
The transition probability has the form
\begin{align}\label{eq:app037}
 P_{0,\mu} \propto \sum_\lambda \left| \sum_\nu \,\langle \Psi_\mu | h_\lambda | \Psi_\nu \rangle\langle \Psi_\nu | \vecs{p} | \Psi_0\rangle \,\right|^2 \,.
\end{align}
As we are primarily interested in the transition that is facilitated by the $\Gamma_3^-$ phonon, we restrict the sum over $\lambda$ to the constituents of this respective phonon branch. The $\Gamma_3^-$ phonon can theoretically scatter into $\Gamma_4^-\otimes\Gamma_3^- =\Gamma_4^+\oplus\Gamma_5^+$ states. For the yellow series only the $\Gamma_5^+$ ortho-exciton states contribute, the green series exhibits $\Gamma_5^+$ ortho- as well as $\Gamma_4^+$ para-exciton states. However, since the $\Gamma_3^-$ phonon transition cannot inflict a change to the spin-configuration of the intermediate state, the scattering into $\Gamma_4^+$ states is not occurring. This can also readily be seen when the coupling strengths of the transitions are evaluated, where the $\Gamma_4^+$ participating states cancel each other out. The $\Gamma_3^-$ phonon transition operator $h_3$\footnote{The momentum subscript $\vecs{Q}$ is dropped here, since it carries no relevance in these considerations.} has two constituents, $\eta_{3_1}$ and $\eta_{3_2}$. Their coupling between the $\Gamma_8^-$ and $\Gamma_6^+$ conduction band can be expressed via
\begin{align}\label{eq:app038a}
 \eta_{3_1} \begin{pmatrix}
             |\mrm{8c},\scst{-\frac{3}{2}}\rangle \\
             |\mrm{8c},\scst{-\frac{1}{2}}\rangle \\
             |\mrm{8c},\scst{+\frac{1}{2}}\rangle \\
             |\mrm{8c},\scst{+\frac{3}{2}}\rangle 
            \end{pmatrix}
         &= \frac{\tilde{D}_{3;68}}{\sqrt{2}}
	    \begin{pmatrix}
                0 \\
                |\mrm{6c},\scst{-\frac{1}{2}}\rangle \\
		-|\mrm{6c},\scst{+\frac{1}{2}}\rangle \\
		0
            \end{pmatrix} \,,
\end{align}
\begin{align}
            \label{eq:app028b}
 \eta_{3_2} \begin{pmatrix}
             |\mrm{8c},\scst{-\frac{3}{2}}\rangle \\
             |\mrm{8c},\scst{-\frac{1}{2}}\rangle \\
             |\mrm{8c},\scst{+\frac{1}{2}}\rangle \\
             |\mrm{8c},\scst{+\frac{3}{2}}\rangle 
            \end{pmatrix}
         &= \frac{\tilde{D}_{3;68}}{\sqrt{2}}
	    \begin{pmatrix}
               - |\mrm{6c},\scst{+\frac{1}{2}}\rangle \\
                0 \\
                0 \\
		|\mrm{6c},\scst{-\frac{1}{2}}\rangle 
            \end{pmatrix}    \,,         
\end{align}
with $\tilde{D}_{\lambda,ij} = \hbar D_{\lambda,ij}\, /\sqrt{2\Omega \rho \,E_\lambda}$. The transformed intermediate $\Gamma_4^-$ states receive the structure of their $\Gamma_5^+$ counterparts, and utilising the orthonormality of the exciton states then eliminates coupling to most of the states. The phonon transition elements then read as follows
\begin{align}
 _\mrm{y\!}\langle XY| \,\eta_{3_1}\, | Z\rangle_\mrm{b}\; &= 0 \label{eq:app29a}\,, \\
 _\mrm{y\!}\langle XY| \,\eta_{3_2}\, | Z\rangle_\mrm{b}\; &= -\frac{\tilde{D}_{3;68}}{\sqrt{2}}\label{eq:app29b} \,,
\end{align}
\begin{align}
_\mrm{g\!}\langle XY| \,\eta_{3_1}\, | Z\rangle_\mrm{v_1} &= 0 \label{eq:app29c} \,, \\
 _\mrm{g\!}\langle XY| \,\eta_{3_2}\, | Z\rangle_\mrm{v_1} &= - \frac{3}{2}\frac{\tilde{D}_{3;68}}{\sqrt{5}}\label{eq:app29d} \,,\\
 _\mrm{g\!}\langle XY| \,\eta_{3_1}\, | Z\rangle_\mrm{v_2} &= 0 \label{eq:app29e}\,, \\
 _\mrm{g\!}\langle XY| \,\eta_{3_2}\, | Z\rangle_\mrm{v_2} &= -\frac{1}{2}\frac{\tilde{D}_{3;68}}{\sqrt{2}}\label{eq:app29f} \,,
\end{align}
In this case, the choice of our intermediate states spares us from a separate evaluation of the $\eta_{3_1}$ component. The transition probability for the phonon assisted transition into the yellow series is then given by
\begin{align}\label{eq:app030}
 P_\mrm{0,y} \propto \left| \,_\mrm{y\!}\langle XY| \,\eta_{3_2}\, | Z\rangle_\mrm{b} \,\,_\mrm{b\!}\langle Z|\,\vecs{p}\,| 0\rangle\,\right|^2 = \frac{1}{3} \tilde{D}^2_{3;68}\,p^2_{35} \,,
\end{align}
the transition probability into the green series results in
\begin{align}\label{eq:app31}
 P_\mrm{0,g} \propto \left|\sum_{i=1}^2 \,_\mrm{g\!}\langle XY| \,\eta_{3_2}\, | Z\rangle_{\mrm{v}_i} \,\,_{\mrm{v}_i\!}\langle Z|\,\vecs{p}\,| 0\rangle\,\right|^2 = \frac{2}{3} \tilde{D}^2_{3;68}\,p^2_{35} \,.
\end{align}
From this we concur, that the ratio between yellow and green $\Gamma_3^-$ phonon-assisted absorption has to be
\begin{align}\label{eq:app032}
 \alpha_\mrm{g}^{\Gamma_3^-}:\alpha_\mrm{y}^{\Gamma_3^-} \;=\; 2 : 1 \,.
\end{align}

\bibliographystyle{apsrev4-1}
\bibliography{abs-citations}{}

\end{document}